\documentclass[aps,pre,twocolumn,showpacs,superscriptaddress, nofootinbib]{revtex4-1}
\usepackage{amsmath,amssymb}
\usepackage{graphics,graphicx,color}
\usepackage{dcolumn,bm}
\usepackage{hyperref}
\usepackage{empheq}
\usepackage{amsfonts}
\usepackage{amsthm}
\usepackage{amssymb}
\usepackage{soul}
\usepackage{url}
\usepackage{float}
\usepackage{subcaption}

\graphicspath{ {./figures/} }

\newcommand{\ncd}{\newcommand}
\ncd{\mrm}    {\mathrm}
\ncd{\beq} {\begin{equation}}
  \ncd{\eeq} {\end{equation}}

\newcommand{\gaug}
{\affiliation{Institute for Theoretical Physics, Georg-August-Universit\"at G\"ottingen, 37077 G\"ottingen, Germany}}

\begin{document}
\title{ Multifractality and statistical localization in highly heterogeneous random networks}

\author{Diego Tapias} 
\email[Email: ]{diego.tapias@theorie.physik.uni-goettingen.de}
\gaug

\author{Peter Sollich}%
\email[Email: ]{peter.sollich@uni-goettingen.de}
\gaug
\affiliation{Department of Mathematics, King's College London, London WC2R 2LS, UK}


\begin{abstract}
 {We consider highly heterogeneous random networks with symmetric interactions in the limit of high connectivity. A key feature of this system is that the spectral density of the corresponding ensemble exhibits a divergence within the bulk. We study the structure of the eigenvectors associated with this divergence and find that they are multifractal with the statistics of eigenvector elements matching those of the resolvent entries. The corresponding localization mechanism relies on the statistical properties of the nodes rather than on any spatial structure around a localization centre. This ``statistical localization'' mechanism is potentially relevant for explaining localization in different models that display singularities in the bulk of the spectrum of eigenvalues.}
\end{abstract}

\maketitle

\section{Introduction}

After three decades of an extensive analysis of its main properties, the configuration model has stood its ground as the simplest but realistic theoretical model for complex networks~\cite{bollobas1998random, newman2018networks}. A particularly appealing property of this model is that the degree distribution can be freely specified, making it possible to capture a large range of random networks varying from highly homogeneous (all nodes with same degree) to very heterogeneous~\cite{metz2020spectral}.

In addition to the topological structure of the network, the edges are often weighted, physically representing the strength of an interaction between nodes~\cite{newman2004analysis, barrat2004architecture}. This is encoded in a matrix that contains the weights of all edges in the network. For the theoretical approach, rather than focusing on individual matrices, one reduces them to their defining properties and represents these in a statistical ensemble of weight matrices~\cite{albert2002statistical}. The interplay between interaction and structure can be studied by analyzing, for instance, the spectral properties of the corresponding ensemble of random matrices (see for instance ref.~\cite{kuhn2008spectra} for a classical work in the case of symmetric sparse matrices or ref.~\cite{metz2019spectral} for a more recent perspective on non--Hermitian random matrices). 

{Analytical methods for computing spectra of configuration model networks can be found in references~\cite{dorogovtsev2003spectra, rogers2008cavity, newman2019spectra, susca2021cavity}. Most of the theory is based on message--passing algorithms and the corresponding cavity (or resolvent) equations~\cite{mezard2009information, susca2021cavity}. Recently, in references~\cite{metz2020spectral, silva2022analytic} an analysis of these equations for  highly heterogeneous networks, i.e.\ those with non-vanishing relative variance of the degree distribution (in the high connectivity limit), has been performed. The main result is that in the thermodynamic limit, classical results of Random Matrix Theory do not apply.} This includes in particular Wigner's semicircle law to describe the spectrum of eigenvalues. Deviations from this are accompanied also by other striking properties, such as the appearance of singularities in the bulk of the spectral density. 


In this paper, we study {highly heterogeneous random networks using} the negative binomial distribution as the degree distribution for the configuration model, with interactions sampled from a Gaussian with zero mean and a variance that scales inversely with the mean connectivity. As has been shown in reference~\cite{silva2022analytic}, this setting leads to a family of networks controlled by a single parameter that allows one to cover the range from highly heterogeneous all the way to essentially homogeneous networks. In reference~\cite{silva2022analytic}, the resolvent equations were derived and solved for this system. Their solution revealed that in the case of highly heterogeneous graphs, there is a divergence in the spectral density at eigenvalue zero. This can be traced back to the distribution of the local Density of States, which acquires a power--law tail leading to an infinite first moment.

{Here we investigate the structure of the eigenvectors} associated with the divergence of the spectral density. Our main finding is that they are localized and exhibit ``strong multifractal'' behavior. We rationalize this by {finding an anticorrelation between the degree of the node and the corresponding amplitude of the eigenvector. It turns out that localization occurs in nodes with low degree (relative to the mean), which typically are far apart from each other.} We denote the corresponding localization mechanism ``statistical localization'' as it is driven by the statistical properties of the node(s). This contrasts with the standard case of Anderson localization on random networks~\cite{kravtsov2018non, tikhonov2021anderson, garcia2022critical}, where eigenvectors are localized on nodes {around a single localization centre}. We conjecture that the new mechanism of statistical localization is relevant also for other types of models that exhibit singularities within the bulk of the spectral density {such as the Poisson random graph with Gaussian couplings~\cite{kuhn2008spectra} or the sparse Barrat--M\'ezard trap model~\cite{tapias2020entropic, tapias2022localization}. }

\section{Heterogeneous weighted random networks}

We provide in this section the essential pieces of information that define the ensemble of random networks that we study. For more details, we refer to the reader to the original paper~\cite{silva2022analytic}. Let us consider a simple and undirected network with $N$ nodes. We consider the configuration model with a negative binomial distribution of degrees $k$,
\begin{align}
  p_k = \frac{\Gamma(\gamma + k)}{\Gamma(\gamma)} \frac{1}{k!} \left( \frac{c}{\gamma} \right)^k \frac{1}{\left( 1+ {c}/{\gamma} \right)^{\gamma + k}}
  \label{negative_binomial}
\end{align}
where $c$ is the mean connectivity, $\Gamma(\cdot)$ is the Gamma function and $0 < \gamma < \infty$ controls the heterogeneity of the distribution~\footnote{In ref.~\cite{silva2022analytic}, the parameter $\gamma$ is denoted by $\alpha$. We use a different convention here because we will need $\alpha$ later in the notation for the spectrum of fractal dimensions.}. Indeed, the relative variance of the degree distribution is given by
\begin{align}
  \frac{ \sigma^2}{c^2} = \frac{1}{c} + \frac{1}{\gamma}
  \label{relvar}
\end{align}
Each node $i$ is assigned a degree $k_i$ drawn randomly from $p_k$, and nodes are then randomly connected following the standard configuration model prescription~\cite{newman2018networks}. If there is a link between nodes $i$ and $j$ we set the random variable $c_{ij} = 1$, otherwise it is set to zero. Additionally we consider interactions $J_{ij}$ between nodes as random variables sampled independently on each link (with $J_{ij}=J_{ji}$) from a distribution $p_J$ with mean zero and standard deviation $J/\sqrt{c}$. Overall, the weight matrix ${\bm{A}}$ has elements
\begin{align}
  A_{ij} = c_{ij} J_{ij}
  \label{weight}
\end{align} 

In the high--connectivity limit $c \to \infty$, it makes sense to consider the distribution of the rescaled degrees $\kappa = k/c$. This distribution $\nu(\kappa)$ is formally defined as follows
\begin{align}
  \nu(\kappa) = \lim_{c \to \infty} \sum_{k =0}^\infty p_k \,\delta \!\left( \kappa - \frac{k}{c} \right)
\end{align}
For $p_k$ as given in eq.~\eqref{negative_binomial}, one finds that $\nu(\kappa)$ is a Gamma distribution with shape parameter $\gamma$ and scale parameter $1/\gamma$, i.e.
\begin{align}
  \nu(\kappa) = \frac{\gamma^\gamma \kappa^{\gamma - 1} {\rm{e}^{-\gamma \kappa}}}{\Gamma(\gamma)}
  \label{negbi}
\end{align}
The regime $\gamma < 1$ characterizes random networks with strongly heterogeneous degrees {as the relative variance of the degree distribution is greater than 1 (cf.\ eq.~\eqref{relvar}).} In the limit $N \to \infty$ ({taken before} $c \to \infty$),  the distribution of eigenvalues, $\rho(\lambda) = \lim_{N\to \infty}\frac{1}{N} \sum_{i=1}^N \delta(\lambda - \lambda_i)$ , of the weight matrix (eq.~\eqref{weight}) exhibits a power law divergence
\begin{align}
  \rho(\lambda) \sim |\lambda|^{\gamma - 1}\, ,  \quad |\lambda| \to 0
  \label{divergence}
\end{align}
with $0 < \gamma < 1$. This behavior can be rationalized via the connection between the resolvent and the spectral density~\cite{potters2020first}, namely
\begin{align}
  \rho(\lambda) =  \lim_{\epsilon \to 0} \lim_{N\to \infty} \frac{1}{\pi N} \sum_{i = 1}^N {\rm{Im}}\, G_{ii} (\lambda -i \epsilon)
  \label{residentity}
\end{align}
where the $G_{ii}$ are the diagonal elements of the resolvent matrix ${\bm{G}} = ((\lambda - i \epsilon) {\bm{I}} - {\bm{A}})^{-1}$, {with $\bm{I}$ the identity matrix}. As a matter of fact, networks generated by the configuration model exhibit locally a tree--like structure~\cite{newman2018networks}, this implies that we can use the cavity method~\cite{rogers2008cavity, susca2021cavity} to estimate the diagonal elements of the resolvent matrix, also called \emph{local Density of States} (lDOS). Ref.~\cite{silva2022analytic} shows that at $\lambda = 0$ the distribution of  $y_i = {\rm{Im}}\,G_{ii}$ in  the limit $\epsilon \to 0$ is given by
  \begin{align}
    P_0(y) = \frac{\gamma^\gamma}{\Gamma(\gamma) J^\gamma} \frac{\exp\left( - \frac{\gamma}{J y} \right)}{y^{\gamma + 1}}
    \label{ldos}
  \end{align}
  By virtue of the resolvent identity (eq.~\eqref{residentity}), the mean value of the lDOS distribution determines the spectral density. From the power-law tail of the distribution~\eqref{ldos} one can then clearly see the origin of the divergence in equation~\eqref{divergence}.

A final result that we want to introduce here comes from reference~\cite{dembo2021empirical}. This states that in the limit $c \to \infty$, the spectral density of the weight matrix $\bm{A}$ is given by the free multiplicative convolution~\cite{speicher2009free, akemann2011oxford} of $\nu(\kappa)$ with the Wigner semicircle law $\rho_{\rm{W}}(\lambda)$. As an implication, one has that the weight matrix can be decomposed as the product of two different matrices, $\bm{X}$ and $\bm{D}$, where the latter is the degree matrix with elements $D_{ij} = \kappa_i \delta_{ij}$ and the former is a random (symmetric) matrix with zeros on the diagonal and off--diagonal elements drawn from a Gaussian distribution with mean zero and variance $J^2/N$. After symmetrization, one can write
  \begin{align}
    {\bm{A}} = {\bm{D}}^{1/2}  {\bm{X}}  {\bm{D}}^{1/2}
    \label{adja}
  \end{align}
 This decomposition is used in reference~\cite{silva2022analytic} to compare the predictions {(in particular, for the spectral density)} from theory to numerical results using exact diagonalization, without explicitly constructing the configuration model. We will use the same approach to obtain the eigenvectors associated with the modes around $\lambda = 0$ numerically.
  
\section{Wavefunction statistics}

\begin{figure}
        \centering
        \begin{subfigure}[b]{0.48\textwidth}
            \centering
            \includegraphics[width=\textwidth]{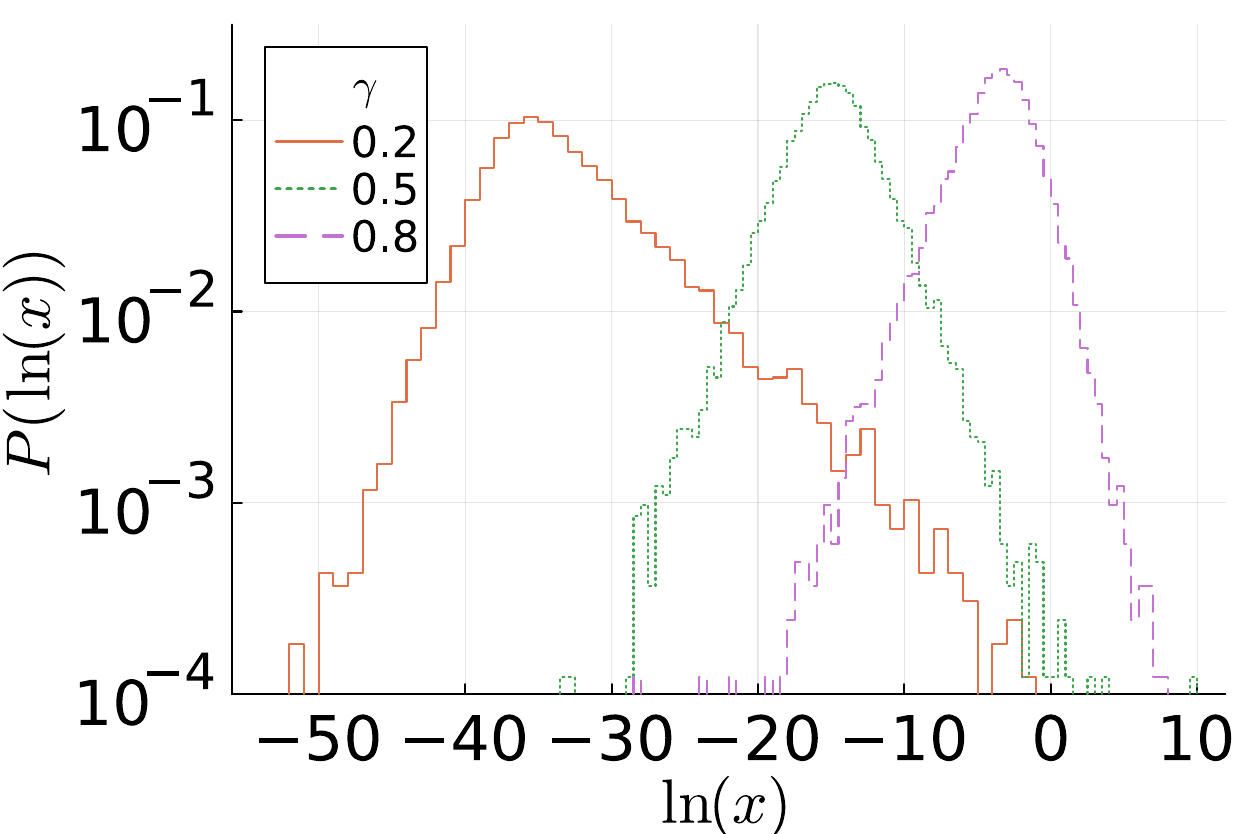}
        \end{subfigure}
        \begin{subfigure}[b]{0.48\textwidth}
            \centering
            \includegraphics[width=\textwidth]{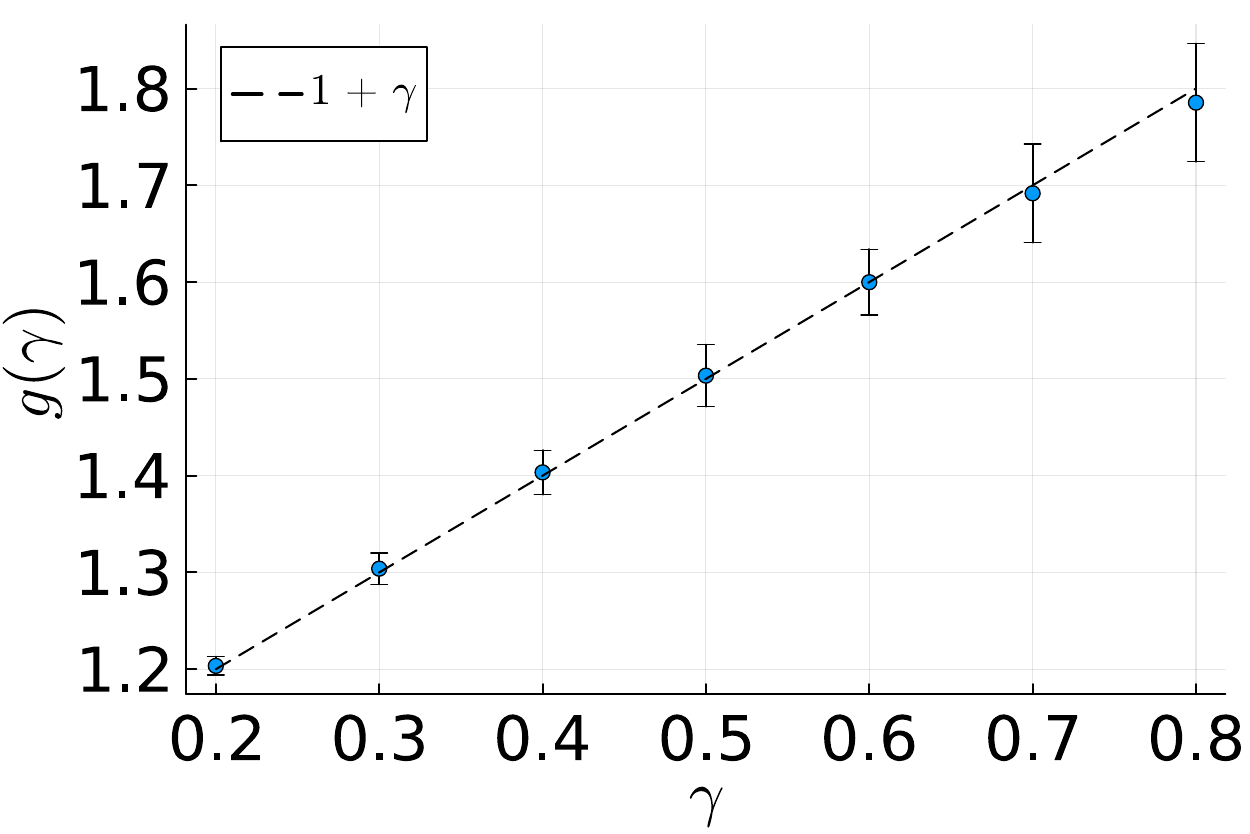}
        \end{subfigure}
        \caption{(a) Distribution of the (logarithm) of squared wavefunction amplitudes, $x_i = N|\psi_i|^2$, for three different eigenvectors of interaction matrices of size $N = 2^{14}$ generated for different heterogeneity parameters $\gamma$. (b) Estimation of the power law tail of the distribution $P(x) \sim x^{-g(\gamma)} $, fitted from data for $x > x_{*}$ with {$x_{*} = 20\, x^{\rm{typ}}$} well above the mode ($\sim x^{\rm{typ}}$) of the distribution.  Error bars show statistics over 128 instances of $\bm{A}$, using the 50 eigenvectors with eigenvalues closest to $\lambda = 0$ for each instance.}
         \label{tail}
       \end{figure}

We exploit the decomposition in equation~\eqref{adja} to generate an interaction matrix $\bm{A}$. Then we diagonalize it with the Lanczos algorithm to extract the eigenvectors with associated eigenvalues closest to zero~\footnote{For this purpose we use an implementation of the \emph{Arpack} package for the \emph{Julia} programming language~\cite{arpack}.}. In Figure~\ref{tail}((a)) we show the distribution of the (scaled) squared eigenvector entries $x_i = N|\psi_i|^2$, which in quantum mechanical language are squared wavefunction amplitudes. The data are obtained from independent instances of $\bm{A}$ for three different values of $\gamma$. The figure suggests that the distributions have power-law tails for large $x$, with $\gamma$-dependent exponents~\footnote{{The figure also suggests the existence of a left power-law tail. This would contribute to the scaling of the exponents $\tau(q)$ for negative $q$ (see eq.~\eqref{legendre}) and equivalently to the multifractal spectrum for $\gamma \alpha > 1$ (cf.\ eq.~\eqref{mfspectrum}). As this piece is not relevant for the understanding of the localization properties, we do not consider it further in our analysis.}}. In fact, we found that the exponents agree with those  for the resolvent, i.e.\ $P(x) \sim P_0(x) \sim x^{-(1 + \gamma)}$ for sufficiently large $x$, as shown in Figure~\ref{tail}((b)). {This agreement of the statistics of the resolvent and eigenvector entries is consistent with previous results for eigenvectors with multifractal properties, see for instance~\cite{monthus2017statistical}.}
       
\begin{figure}
            \centering
            \includegraphics[width=0.48\textwidth]{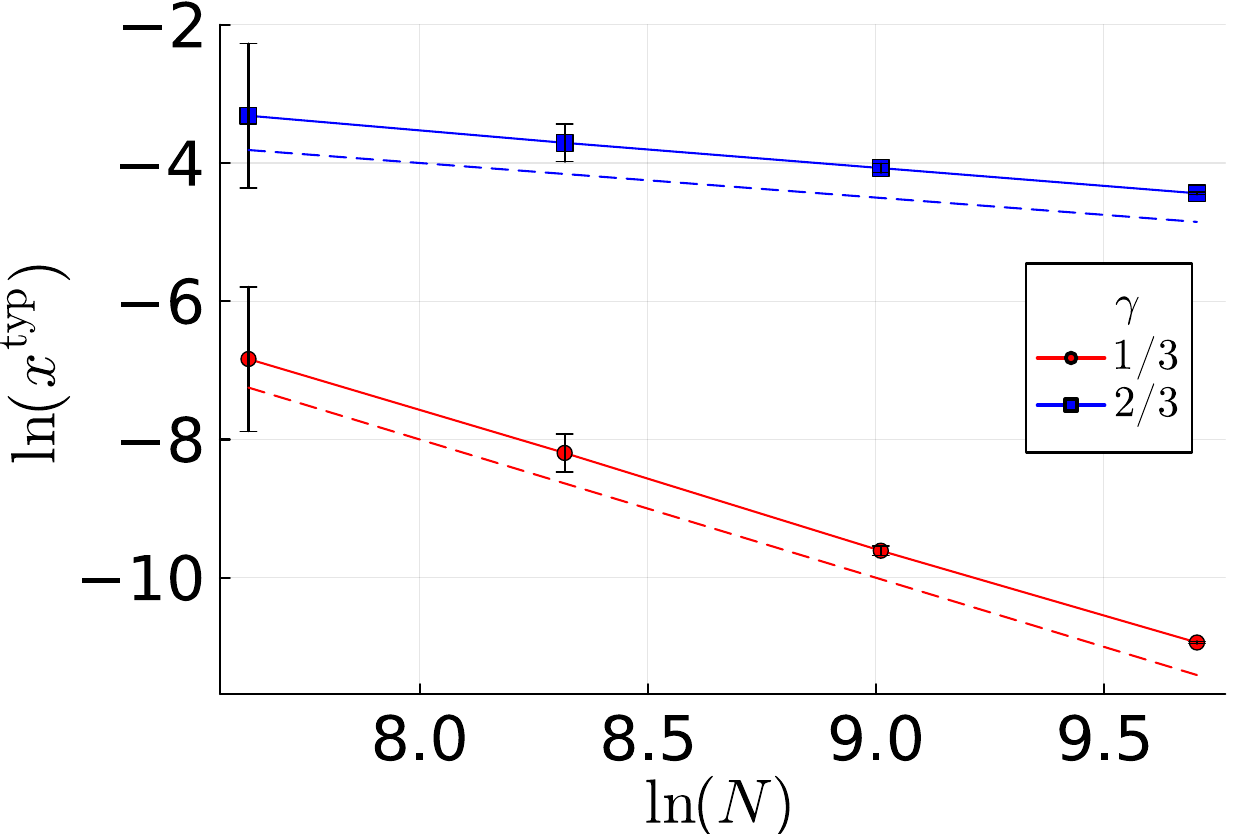}
            \caption{Scaling of typical value defined as $x^{\rm{typ}} = {\rm{e}}^{\langle \ln x \rangle}$ with system size for $\gamma \in \{1/3, 2/3\}$. The prediction $x^{\rm{typ}} \sim  N^{1 - 1/\gamma}$  is shown by the dashed lines.}
            \label{typ_scaling}
          \end{figure}

\begin{figure}
            \centering
            \includegraphics[width=0.48\textwidth]{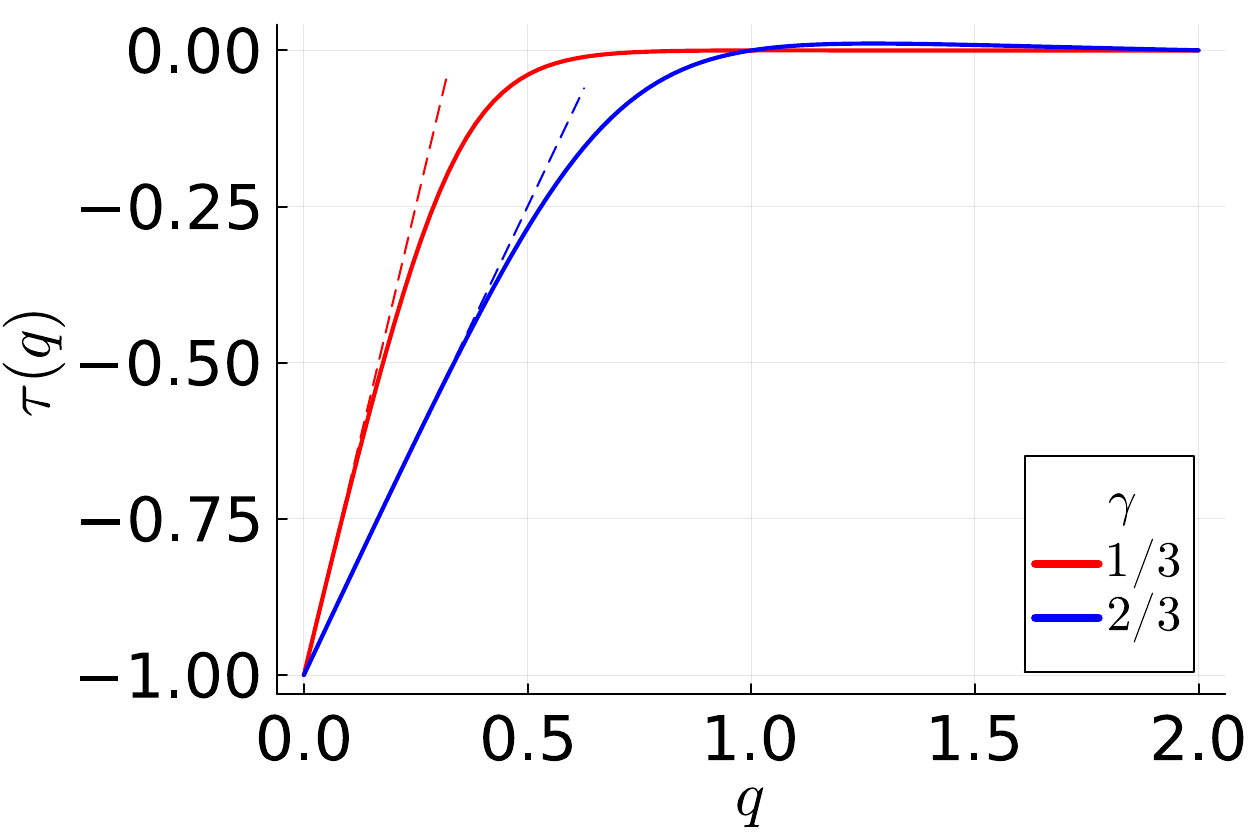}
        \caption{(a) Scaling of the wavefunction moments as encoded in the exponents $\tau(q)$ for $N = 2^{14}$. The dashed lines correspond to the linear piece of the theoretical prediction~\eqref{exponents}. See numerical details in the appendix.}
        \label{mf_nb} 
      \end{figure}

The power-law tail motivates the following \emph{ansatz} for the whole distribution above a scale $x_{\min}(N)$,  close to the mode of the distribution
      \begin{align}
        P(x)  = b(N) x^{-(\gamma + 1)}
        \label{ansatz}
      \end{align} 
      with $b(N)$ and $x_{\min}(N)$ functions to be determined. Their scaling can be estimated by using the normalization of $P(x)$ and its first moment~\cite{monthus2017statistical}; the latter follows from the normalization of the eigenvectors, $1=\sum_i |\psi_i|^2 = N^{-1}\sum_i x_i$. The first condition, namely, $\int P(x)\,dx = 1$ gives
      \begin{align}
        1 \approx \int_{x_{\min}}^N P(x)\,dx  \approx \gamma b x_{\min}^{-\gamma} 
      \end{align}
      whereas the second one, namely $\int xP(x)\,dx = 1$, yields
           \begin{align}
       1 \approx \int_{x_{\min}}^N x P(x)\,dx  \approx \gamma b N^{-\gamma + 1} 
           \end{align}
           These two relations imply $b \sim N^{\gamma - 1}$ and $x_{\min}  \sim N^{1 - 1/\gamma}$.  As $x_{\min} $ sets the scale from which the power--law becomes valid, we identify this as the typical value of the distribution. This scaling is confirmed using results from Exact Diagonalization as can be appreciated in Figure~\ref{typ_scaling}.
          Substituting the scaling for $b(N)$ in our ansatz (eq.~\eqref{ansatz}), we write the distribution $P(x)$ in the generic way
          \begin{align}
            P(x) = \frac{A}{x} N^{\gamma - 1} x^{- \gamma}  
        \label{px}
          \end{align}
          with $A = O(N^0)$ a normalization constant. The spectrum of fractal dimensions $f(\alpha)$, which is defined 
as~\cite{de2014anderson, kravtsov2015random}
          \begin{align}
            f(\alpha) = \lim_{N \to \infty} \ln \left( xN P(x) \right) /\ln N
            \label{falpha}
          \end{align}
          where on the r.h.s.\ $x=N^{1-\alpha}$ or conversely $\alpha = 1 - \ln x/\ln N$. Substitution of eq.~\eqref{px} into eq.~\eqref{falpha} gives the expression
           \begin{align}
             f(\alpha) = \gamma \alpha \, , \quad {\rm{for}} \quad \gamma \alpha \leq 1
             \label{mfspectrum}
           \end{align}
 {where the upper cutoff $\alpha_{\max} = \frac{1}{\gamma}$ corresponds to the lower cutoff $x_{\min}$.}  Finally, the Legendre transform of equation~\eqref{falpha} gives the set of exponents characterizing the $q$--th moment of the distribution 
           $I_q(N) = \sum_i |\psi_i|^{2q}  \propto N^{-\tau(q)}$
The Legendre transform relation reads explicitly~\cite{monthus2017statistical}
           \begin{align}
             -\tau(q) = \max_\alpha\,[f(\alpha) - q\alpha]
             \label{legendre}
           \end{align}
           Substitution of $f(\alpha)$ (eq.~\eqref{mfspectrum}) into equation~\eqref{legendre} yields
           \begin{align}
             \tau(q) =
             \begin{cases}
               \frac{q}{\gamma} - 1 \qquad &q \leq \gamma \\
               0 \qquad &q > \gamma
             \end{cases}
             \label{exponents}
           \end{align}
          Figure~\ref{mf_nb} compares this result with Exact Diagonalization.

            \section{Discussion}

            The scaling of the moments $I_q$ as given by equation~\eqref{exponents}  has been found across different models and the corresponding phase has been described {variously} as ``quasilocalized''~\cite{kravtsov2015random}, or localized with multifractal properties~\cite{monthus2016localization}, or localized with ``strong multifractal'' behavior~\cite{garcia2022critical}. The corresponding models are the Gaussian Rosenzweig--Porter, random Levy matrices, and the Anderson model on small--world networks, respectively. {References~\cite{kravtsov2015random, monthus2016localization} point to a power--law (instead of exponential) localization {as} the origin of this kind of behavior. In contrast, reference~\cite{garcia2022critical} (see also~\cite{garcia2020two}) focusses on the existence of a length scale that characterizes the decay of wave functions along the typical branches as responsible for the scaling above. These mechanisms, however, do not seem to be operative for our system. }

\begin{figure}
        \centering
        \includegraphics[width=0.48\textwidth]{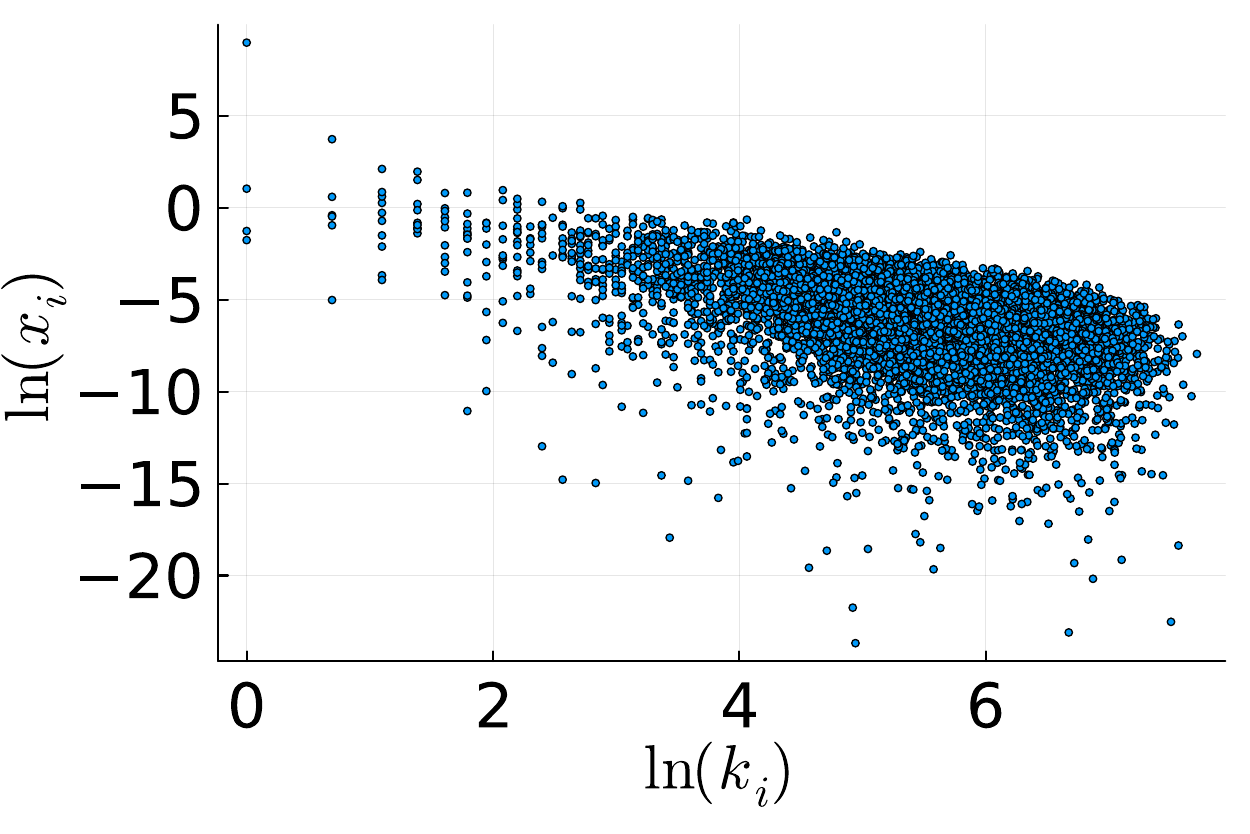}
                \caption{Correlation among the {scaled} eigenvector mass, $x_i = N|\psi_i|^2$ and the node degree $k_i$ for the eigenvector with eigenvalue closest to $0$ for an instance constructed via equation~\eqref{weight}, for a finite configuration model network with mean connectivity $c = 400$ and $\gamma = 1.2$ {and of size $N=8187$.  (This network is the giant connected component of the original configuration model graph with $2^{13}=8192$ nodes.)}}
        \label{scatter_nb}
      \end{figure}

            In order to find the mechanism that governs the behavior of {our}  model, we construct finite instances of the networks generated with the configuration model. Then we investigate the correlation between the squared eigenvector entry (or ``mass'') and the degree of the node. Our finding is that low--degree nodes, typically leaves (i.e. nodes with degree one), concentrate the mass of the eigenvectors. For mildly heterogenous networks, i.e. those with $1 < \gamma < 2$, the anticorrelation between degree and eigenvector mass is easily visible in a scatter plot (c.f.~Figure~\ref{scatter_nb}). If we extrapolate this finding to $c \to \infty$, we expect that for a given instance nodes with small relative degree will concentrate the mass of the eigenvector, and what is more relevant, that those nodes may be far apart from each other on the network. {Indeed, we find that the dominant nodes from figure~\ref{scatter_nb} lie at distances of the order of the graph diameter from each other.} Thus what matters for localization is the identity of each node and its statistical properties, rather than its spatial location in the network. In this sense, we say that the system exhibits ``statistical localization'' and that this leads to the multifractal behavior encoded in the exponents~\eqref{exponents}.

            To  illustrate further the mechanism described above, consider the Bouchad trap model on a sparse random graph~\cite{margiotta2018spectral}. The ground state of this system is the Boltzmann distribution, and the set of exponents characterizing the scaling of the moments is of the form~\eqref{exponents}, with  temperature instead of $\gamma$ as the control parameter (see equation 4.15 in reference~\cite{riccardothesis}). Thus, for finite instances, the nodes that would carry most of the mass of the vector are the ones with the largest energies, corresponding to the deepest traps. Those nodes are spatially uncorrelated and are  {instead identified by local statistical properties}.

         {At this point, it is worth comparing our results with earlier observations of localization in heterogenous networks. As a representative example of this class we consider the Laplacian on Poisson (Erd\"os--R\'enyi) random graphs~\cite{biroli1999single, kuhn2008spectra} (see also~\cite{monasson1999diffusion}). In both cases, for small $|\lambda|$, localization is driven by low degree nodes. However, for our model, this happens in the bulk of the spectrum and is accompanied by a divergence in the spectral density; whereas for the Poisson random graph, localization happens in the (Lifshitz~\cite{khorunzhiy2006lifshitz}) tail of the spectrum, which is separated from the bulk by a mobility edge. Additionally, while it has been observed that localized states in this tail are centred on geometric defects with abnormally low connectity~\cite{biroli1999single} and that the density of states is dominated by low degree nodes~\cite{kuhn2008spectra}, it remains an open question whether those states exhibit multifractal behavior and the existence of multiple localization centres. This would be interesting to study in future work.
 }
      
      In conclusion, we have analysed the wavefunction statistics for highly heterogeneous random graphs from the configuration model defined by a negative binomial degree distribution and in the limit of high connectivity. We have found that the distribution of the eigenvector mass for the modes that contribute to the divergence of the spectral density is a power-law, with the same exponent as the one for the local Density of States. Those eigenvectors exhibit strong multifractality, and for finite graphs are essentially localized in the lowest-degree nodes. On this basis, we introduce the concept of ``statistical localization'' that is to be contrasted with the standard one of ``spatial localization''.

      We leave for future work an extensive analysis of the eigenstate correlations of these modes as has been done for the Anderson model~\cite{tikhonov2021eigenstate} or the Gaussian Rosenzweig-Porter model~\cite{kravtsov2015random}, {in order to investigate specific footprints of the statistical localization mechanism}. It will also be interesting to carry out an analysis {similar to the one in this paper} for the sparse Barrat-M\'ezard trap model~\cite{tapias2020entropic, tapias2022localization}, which like the model considered here exhibits divergences in the bulk of the spectral density. {Work in this direction is in progress.} {Finally, we point out that the phase described in earlier studies as a ``frozen phase''  has similar phenomenology to the one observed here~\cite{chamon1996localization, carpentier2001glass, chou2014chalker}, and thus may potentially also be viewed as an instance of the notion of statistical localization that we have introduced here. A quantitative analysis along these lines is left as an interesting task for future work.}

\appendix*
      \section{Numerical details}
    We generate $2^{21}/N$  instances  of weight matrices of size $N \in \{2^{13} , 2^{14}\}$, constructed according to equation~\eqref{adja}. For each instance, we extract the 50 eigenvectors closest to $\lambda = 0$. Then, for each eigenvector $\psi^{(\alpha)}$ with entries $\psi^{(\alpha)}_i$ we compute the $q$--moment (of the squared entries) as $I_q = \sum_{i=1}^N |\psi^{(\alpha)}_i|^{2q}$. Finally, we obtain the typical value over all the vectors as  $I_q^{\rm{typ}} = {\rm{e}}^{\langle \ln I_q \rangle}  \sim N^{-\tau(q)}$  and estimate $\tau(q)$ as:
      \begin{align}
        \tau(q) = - \frac{\ln I_q^{\rm{typ}}(N )  - \ln I_q^{\rm{typ}}(N - \delta N)  }{\ln(\delta N)}
      \end{align}
      with $N = 2^{14}$ and $\delta N=2^{13}$.

\bibliographystyle{unsrt}
\bibliography{sample}
\end{document}